\documentclass[conference]{IEEEtran}

\usepackage{cite}
\usepackage{amsmath,amssymb,amsfonts}
\usepackage{algorithm}
\usepackage{algorithmic}
\usepackage{graphicx}
\usepackage{textcomp}
\usepackage{xcolor}
\usepackage{amsmath,amssymb,amsthm} 
\usepackage{mathtools} 
\usepackage{bm} 

\usepackage{subcaption}
\usepackage{adjustbox}
\usepackage{booktabs}  
\usepackage{siunitx}   
\usepackage{tikz}
\usepackage{caption}
\usepackage{placeins}

\usepackage{hyperref}
\hypersetup{
    colorlinks=true,
    linkcolor=red,
    citecolor=red,
    urlcolor=red
}
\usepackage[capitalise]{cleveref}

\usepackage[
  left=0.680in,
  right=0.680in,
  top=0.75in,
  bottom=0.99in
]{geometry}

\newcommand\iidsim{\stackrel{\text{i.i.d.}}{\sim}}

\newcommand\ourmethod{\texttt{AN-SNIS}}

\def\BibTeX{{\rm B\kern-.05em{\sc i\kern-.025em b}\kern-.08em
    T\kern-.1667em\lower.7ex\hbox{E}\kern-.125emX}}

\begin{document}

\title{Towards Adaptive Self-Normalized Importance Samplers} 
\author{\IEEEauthorblockN{Nicola Branchini}
\IEEEauthorblockA{\textit{School of Mathematics} \\ \emph{Maxwell Institute for Mathematical Sciences} \\
\textit{University of Edinburgh}\\
Edinburgh, United Kingdom} 
\and
\IEEEauthorblockN{Víctor Elvira}
\IEEEauthorblockA{\textit{School of Mathematics} \\ \emph{Maxwell Institute for Mathematical Sciences} \\
\textit{University of Edinburgh}\\
Edinburgh, United Kingdom} 
}

\maketitle

\begin{abstract}
The self-normalized importance sampling (SNIS) estimator is a Monte Carlo estimator widely used to approximate expectations in statistical signal processing and machine learning.

The efficiency of SNIS depends on the choice of proposal, but selecting a good proposal is typically unfeasible. In particular, most of the existing adaptive IS (AIS) literature overlooks the optimal SNIS proposal. 

In this paper, we introduce an AIS framework that uses MCMC to approximate the optimal SNIS proposal within an iterative scheme. This is, to the best of our knowledge, the first AIS framework targeting specifically the SNIS optimal proposal. We find a close connection with ratio importance sampling (RIS), which also brings a new perspective and paves the way for combining techniques from AIS and RIS. We outline possible extensions, connections with existing MCMC-driven AIS algorithms, theoretical directions, and demonstrate performance in numerical examples.

\end{abstract}


\section{Introduction}
Intractable integrals are common in science and engineering, particularly in statistical signal processing and Bayesian statistics when estimating marginal likelihoods and predictive probabilities \cite{doucet2005monte,vehtari2017practical,stoica2022monte}. Monte Carlo methods, primarily importance sampling (IS) and Markov chain Monte Carlo (MCMC), are standard tools for approximating such integrals. In IS, integrals are estimated as weighted averages of samples, which are obtained by simulating from a proposal distribution. The celebrated \emph{self-normalized IS} (SNIS) estimator allows one to estimate integrals even in the presence of unknown normalizing constants. The performance of SNIS hinges on the selection of an effective proposal distribution. Although many works in the adaptive IS (AIS) literature focus on optimizing proposals for unnormalized IS (UIS) \cite{bugallo2017adaptive,llorente2022mcmc,portier2018asymptotic,delyon2021safe,bianchi2024stochastic}, little attention has been paid to adapting a proposal specifically tailored for SNIS. Because SNIS naturally allows one to use proposals that can be evaluated only up to a normalizing constant, it is often combined with MCMC. The main challenge is that the optimal proposal for SNIS, unlike in unnormalized IS (UIS), cannot be evaluated, even up to a normalizing constant. Indeed, a good selection of proposal for SNIS is an active research line  \cite{rainforth2020target,llorente2023target,branchini2024generalizing}.
Similar \emph{doubly intractable} problems are common in the literature, appearing in pseudo-marginal MCMC \cite{andrieu2009pseudo,andrieu2010particle,doucet2015efficient}, noisy MCMC \cite{alquier2016noisy}, and other methods requiring multiple layers of approximation \cite{10.1214/15-STS523}. In parallel, many works also integrate MCMC with SNIS in various ways. For instance, in \cite{schuster2020markov} and \cite{10.1214/20-EJS1680}, different schemes are proposed in order to weight and recycle non-accepted samples in the estimator. In  \cite{vihola2020importance}, the theoretical properties of SNIS estimates with MCMC samples are studied. More sophisticated schemes exist, e.g., combining multiple MCMC chains with multiple IS \cite{roy2018estimating,silva2024robust} or combinations with tempered Gibbs sampling \cite{zanella2019scalable} - all of which could be considered to provide more advanced algorithms within our proposed framework. 

In this work, we present a new Monte Carlo framework that we call \emph{adaptive nested self-normalized importance sampler}, \ourmethod{}, in which the integral of interest via an MCMC chain that iteratively targets a refined approximation of the optimal SNIS proposal. Unlike most existing adaptive IS methods, our approach does not restrict the proposal to a (typically limiting) parametric family.
We find a close connection between the proposed scheme and adaptive versions of ratio IS (RIS) \cite{chen1997monte} (see also \cite[Section 4.1.2.]{llorente2023marginal}). This connection offers a promising direction for bringing insights and schemes from the RIS literature \cite{llorente2023marginal} to develop novel AIS algorithms for the SNIS estimator.

Our approach differs from existing IS methods that combine SNIS and MCMC, as none of them incorporates the optimal SNIS proposal.
We present the performance of an initial, simple instantiation of the proposed framework in a numerical example. While designed to prioritize visualization and intuition, this example demonstrates a significant performance improvement comparable sampling from the optimal UIS proposal.

The rest of the paper is organized as follows. We introduce the problem statement in Section \ref{sec_prob} and review relevant background. In Section \ref{sec:overall}, we present our framework for adaptive IS with the SNIS estimator driven by MCMC, \ourmethod{}. In \cref{sec:theory_connecs}, we discuss connections in the literature, theoretical properties, and future directions. In \cref{sec:experiments} we test \ourmethod{} in a numerical example. Finally, we give some conclusions in \cref{sec:conclusions}.

\section{Problem statement and background}
\subsection{Problem statement}
\label{sec_prob}

Let $\pi(x)$ be a target density of interest on $\mathcal{X} \subset \mathbb{R}^D$ corresponding to measure $\pi(dx)$. We are interested in approximating expectations with respect to test functions $\varphi: \mathcal{X} \to \mathbb{R}$, $\varphi \in \mathcal{L}_{1}(\pi)$, i.e., $\int |\varphi(x)| 
\pi(dx)  < \infty$,
\begin{align}\label{eq:integral}
    \mu = \mathbb{E}_{\pi} \left[ \varphi(x) \right] = \int_{\mathcal{X}} \varphi(x) \pi(x) dx .
\end{align}
We remark that $\varphi$ can be negative, so can $\mu$. This prevents the straightforward application of techniques for estimating ratios of normalizing constants, as we discuss later.

\subsection{Importance sampling}
When only an unnormalized version $\widetilde{\pi}(x)$, i.e., $ \pi(x) = \widetilde{\pi}(x) / Z_{\pi}$ is available, the self-normalized importance sampling (SNIS) estimator \cite{mcbook,doi:https://doi.org/10.1002/9781118445112.stat08284} is widely used to approximate \eqref{eq:integral} as follows:
\begin{align}\label{eq:basic_snis}
    \widehat{\mu}_{\text{SNIS}} =  \sum_{n=1}^N \overline{w}^{(n)} \varphi(x^{(n)}) , \qquad \overline{w}^{(n)} = w^{(n)} / \sum_{n=1}^N w^{(n)} 
\end{align}
where $\{ x^{(n)} \}_{n=1}^N$ are samples from the proposal PDF \( \pi \ll q \), and their importance weights are \(\{w^{(n)} = \widetilde{\pi}(x^{(n)}) / q(x^{(n)})\}_{n=1}^N\). Consistency is ensured also with non-i.i.d. samples from an Harris-ergodic chain \cite{vihola2020importance}. The SNIS estimator, being a ratio, allows the proposal PDF in the denominator to be replaced by its unnormalized version, \(\widetilde{q}(x)\). 

\noindent\textbf{The optimal proposal in SNIS.} The SNIS estimator has an \emph{optimal} proposal that minimizes asymptotic variance \cite{hesterberg1988advances}, 
\begin{align}\label{eq:q_star_snis}
q^{\bigstar}_{\text{SNIS}}(x) &= \frac{\pi(x) |\varphi(x) - \mu|}{\int \pi(x) |\varphi(x) - \mu| dx } = \frac{\widetilde{\pi}(x) |\varphi(x) - \mu|}{\int \widetilde{\pi}(x) |\varphi(x) - \mu| dx } , \nonumber 
\end{align}
see \cite{portier2018asymptotic,branchini2024generalizing} for derivations. This distribution can differ significantly from \(\pi(x)\) or the optimal unnormalized IS proposal, \(\pi(x) |\varphi(x)|\) \cite{lamberti2018double,rainforth2020target,llorente2023target,branchini2024generalizing}, see \cref{sec:experiments}. It is worth remembering that sampling from $\pi$ can be very suboptimal (e.g., in rare events estimation or Bayesian sensitivity analysis). The optimal SNIS proposal is unfortunately almost always overlooked in the AIS and variational inference (VI) literature.

\section{The adaptive nested SNIS framework}\label{sec:overall}
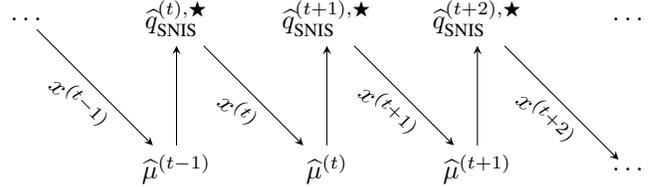
\begin{figure}[h]
\begin{center} 
\begin{tikzpicture}[
  >=stealth,
  ->,
  node distance=2cm, 
  scale=0.9, 
  every node/.style={font=\normalsize}
]

\node (q1) {\(\widehat{q}_{\text{SNIS}}^{(t),\bigstar}\)};
\node (q0) [left of=q1] {\( \dots \)};
\node (q2) [right of=q1] {\(\widehat{q}_{\text{SNIS}}^{(t+1),\bigstar} \)};
\node (q3) [right of=q2] {\(\widehat{q}_{\text{SNIS}}^{(t+2),\bigstar}\)};
\node (qdots) [right of=q3] {\(\dots\)}; 

\node (x1) [below of=q1] {\(\widehat{\mu}^{(t-1)}\)};
\node (x2) [below of=q2] {\(\widehat{\mu}^{(t)}\)};
\node (x3) [below of=q3] {\(\widehat{\mu}^{(t+1)}\)};
\node (xdots) [below of=qdots] {\(\dots\)}; 
\node (nothing) [right of=x3] {\(  \)}; 

\draw (x1) -- (q1);
\draw (x2) -- (q2);
\draw (x3) -- (q3);

\draw (q0) to node[below,sloped]{\(x^{(t-1)}\)} (x1);
\draw (q1) to node[below,sloped]{\(x^{(t)}\)} (x2);
\draw (q2) to node[below,sloped]{\(x^{(t+1)} \)} (x3);
\draw (q3) to node[below,sloped]{\(x^{(t+2)} \)} (nothing);
\end{tikzpicture}
\caption{Illustration of the dependencies within \ourmethod{}.}
\label{illustr}
\end{center}
\end{figure}

We now present our proposed \ourmethod{} framework, that adaptively improves the SNIS proposal. 
The framework is presented in Algorithm \ref{alg:simple-alg-2} and run for $T \in \mathbb{N}$ iterations for adaptation.
At each iteration $t$, an MCMC chain is run for $J \in \mathbb{N}$ steps.
%
We will use a simple random walk Metropolis (RWM) for clarity and concreteness in the exposition. 
We build a SNIS estimator of $\mu$ in Eq. \eqref{eq:current_est} using the samples of the current $t$-th chain, with unnormalized weights given by
\begin{equation}\label{eq:weight}
    w^{(t)}(x) = \frac{\widetilde{\pi}(x)}{\widetilde{\pi}(x)|\varphi(x) - \widehat{\mu}^{(t-1)} |} = (|\varphi(x) - \widehat{\mu}^{(t-1)} |)^{-1} ,
\end{equation}
where $\widehat{\mu}^{(t-1)}$ is the estimate of $\mu$ at previous iteration. In experiments (\cref{sec:experiments}) and for simplicity, our final estimate is $\widehat\mu = \frac{1}{T} \sum_{t=1}^{T} \widehat{\mu}^{(t)}$. Below we discuss more possible options.
\noindent\textbf{Interpretation.} Our framework can be seen as a Markov chain that targets \emph{a sequence of approximations} of the optimal proposal $q_{\text{SNIS}}^{\bigstar}$.
It is possible to show, under typical assumptions for MCMC, that for sufficiently large $J$, the chain obtains samples from $\widehat{q}_{\text{SNIS}}^{(t),\bigstar}$ \cite{vihola2020importance}.
 The method is \emph{nested}, in the sense that, at each iteration $t$, samples following a Markov chain are used to construct an estimate $\mu^{(t)}$ of the integral \eqref{eq:integral}, that \emph{parametrizes} the next proposal $q^{(t+1)} \propto \widetilde{\pi}(x)|\varphi(x) - \widehat{\mu}^{(t)} |$. \cref{illustr} shows an illustration of the dependencies between the generated samples, the estimates of $\mu$, and the approximations of the optimal SNIS proposal $\widetilde{\pi}(x)|\varphi(x) - \widehat{\mu} |$. The importance weight in \cref{eq:weight} reminds of the weight $w(x) = |\varphi(x)|^{-1} $ which arises when using the optimal UIS proposal, as explained in \cite[Chapter~3]{robert1999monte}.

\noindent\textbf{Alternatives within the framework.} We stress that while we present \ourmethod{} with a RWM, other Markov kernels and acceptance functions could be used.  
The estimators $\widehat{\mu}^{(t)}$ could be linearly combined as
    $\widehat{\mu} = \sum_{t=1}^{T} \nu_{t} \widehat{\mu}^{(t)}, $ 
where $ \sum_{t=1}^{T} \nu_{t} = 1$. Beyond equal weighting ($\nu_{t} = 1/T$), options include weights based on effective sample size \cite{elvira2022rethinking,nguyen2014improving}, empirical variance \cite{Cappe08}, using only the final iteration \cite{miller2021rare}, or the square root rule \cite{owen2020square} with $\nu_t \propto t^{1/2}$, though conservative values $0 < 1/k < 1/2$ (emphasizing later iterations) often perform better in practice. The algorithm could be modified to use all past samples when estimating $\mu^{(t-1)}$, similar to adaptive multiple IS \cite{cornuet2012adaptive}. However, we advise against re-normalizing weights across iterations \cite{cornuet2012adaptive,elvira2017improving}, as each iteration targets a different normalizing constant ratio. While our samples are not ``properly weighted'' in the Liu sense \cite{liu2001monte,elvira2019generalized}, this is not necessary for estimating $\mu$ when using SNIS.

\section{Connections in the literature and discussion}\label{sec:theory_connecs}

\subsection{Connections with existing methods}
\textbf{Ratio IS and bridge sampling.} Interestingly, there is a close connection between the proposed framework and the so-called ratio IS (RIS) \cite{chen1997monte} (see a review in \cite{llorente2023marginal}). RIS targets the estimation of a ratio of two normalizing constants $\mu > 0$,  $\mu = \int \widetilde{\pi}_0(x) dx / \int \widetilde{\pi}_1(x) dx = Z_{\pi_0} / Z_{\pi_1}$, where $\widetilde{\pi}_0(x)$ and $\widetilde{\pi}_1(x)$ are unnormalized densities. Then, as in SNIS, an estimator of the ratio is constructed as the ratio of two UIS estimators with the same samples, $\widehat{\mu} = \sum_{n} \widetilde{\pi}_0(x^{(n)}) / q(x^{(n)}) \Big/  \sum_{n} \widetilde{\pi}_1(x^{(n)}) / q(x^{(n)})$. It is easy to see that the proposal minimizing asymptotic variance is the same as SNIS, $\propto |\pi_0(x) - \mu \cdot \pi_1(x)|$ as shown in~\cite{chen1997monte}. Differently from our proposed \ourmethod{}, no iterative RIS scheme targeting the optimal proposal has been proposed (while it is mentioned that this would be possible in~\cite{llorente2023marginal}). The better-known \emph{bridge sampling} (BS) also targets ratios of normalizing constants. However, BS requires sampling from $\pi_1(x)$ and $\pi_0$ (or using MCMC); this is not possible in general for our $\mu = \int \varphi(x) \pi(x) dx$, as $\varphi(x)$ can take negative values. It has also been proved that the asymptotic MSE for RIS is lower than optimal bridge sampling, which means that our \ourmethod{} would also inherity this asymptotic superiority.

\noindent \textbf{Other MCMC-driven AIS samplers.} Our \ourmethod{} connects with layered adaptive IS (LAIS) \cite{martino2017layered}, but unlike existing MCMC-driven samplers \cite{martino2017layered,llorente2022mcmc}, directly targets the optimal SNIS proposal. In LAIS terminology, \cref{illustr} represents an evolving upper layer; AT-LAIS \cite{martino2017anti} uses tempered distributions for different purposes. Other approaches \cite{lamberti2018double,rainforth2020target,branchini2024generalizing} modify adaptation for SNIS but use separate proposals for denominator and numerator instead of targeting the optimal SNIS proposal directly. In \cite{branchini2024generalizing}, proposals are combined with a joint distribution optimized for asymptotic variance. Similarly, \cite{paananen2021implicitly} approximates $q_{\text{SNIS}}^{\bigstar}$ using a \emph{mixture} $\alpha_1 \pi(x) + \alpha_2 \pi(x) |\varphi(x)|$, which crucially differs from the true SNIS optimal proposal that involves \emph{subtraction} rather than addition.

\begin{figure*}[t]
    \centering
    \begin{subfigure}[b]{0.31\textwidth}
        \centering
        \includegraphics[width=0.9\textwidth]{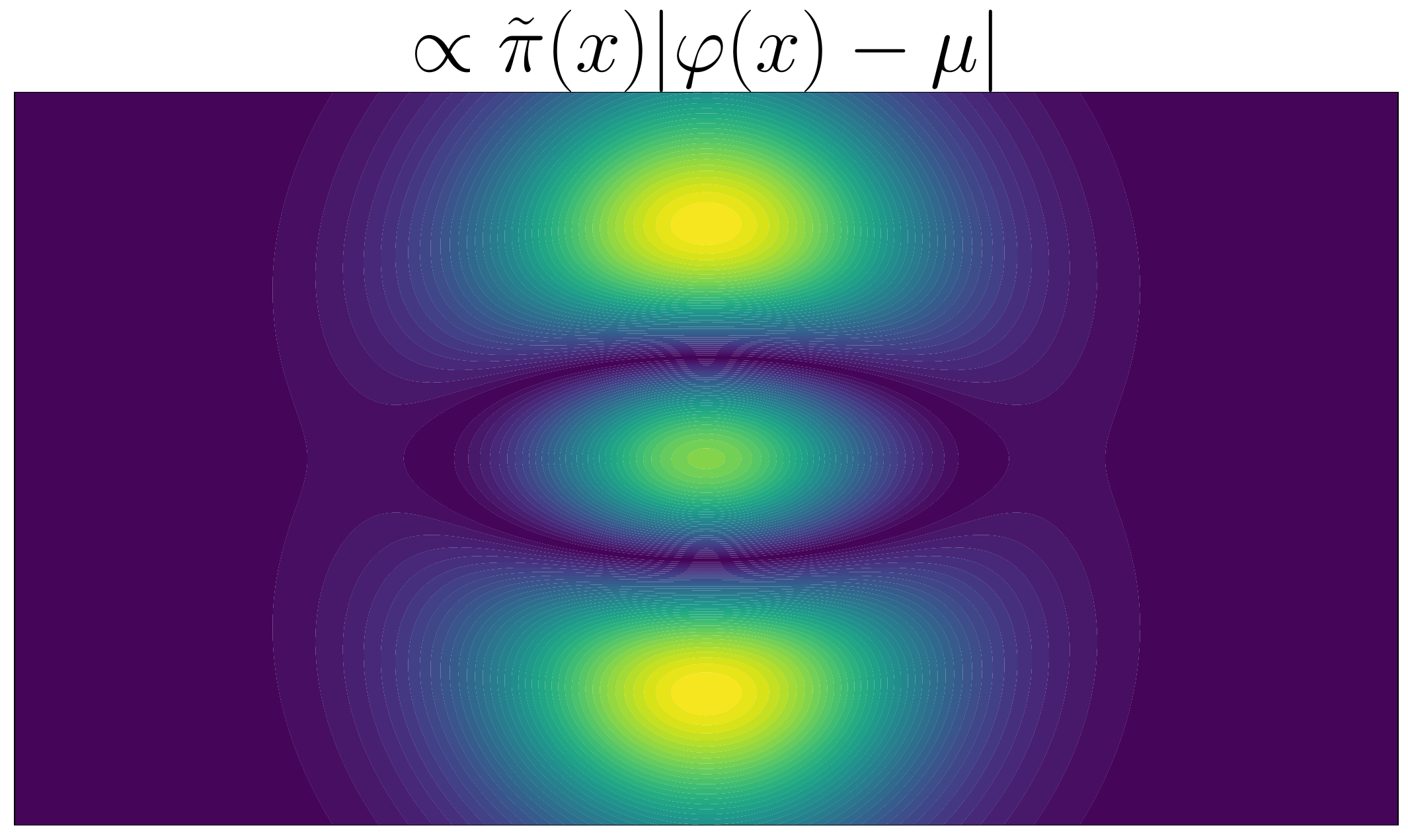}
        \label{fig:plot1}
    \end{subfigure}
    \hfill
    \begin{subfigure}[b]{0.31\textwidth}
        \centering
        \includegraphics[width=0.9\textwidth]{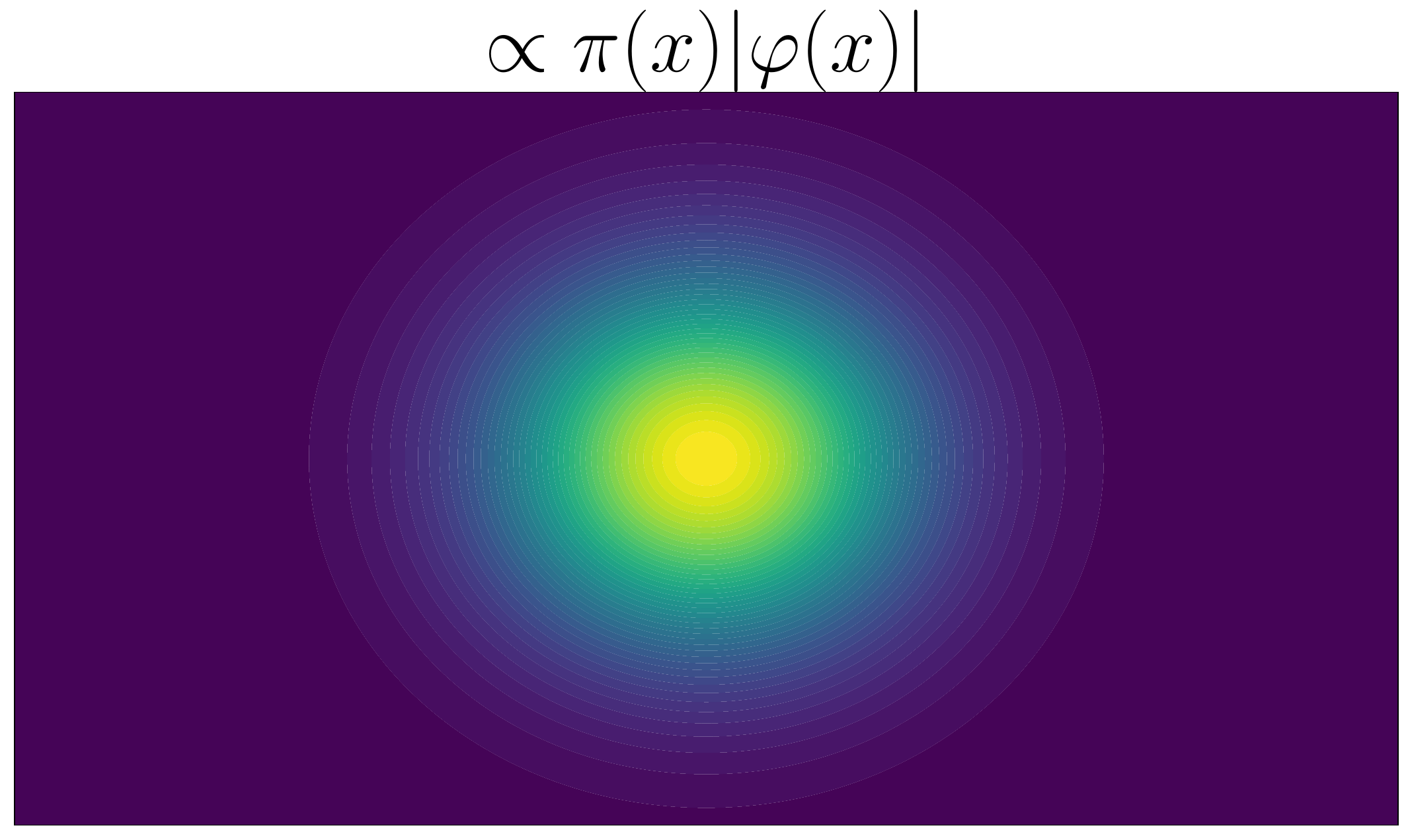}
        \label{fig:plot2}
    \end{subfigure}
    \hfill
    \begin{subfigure}[b]{0.31\textwidth}
        \centering
        \includegraphics[width=0.9\textwidth]{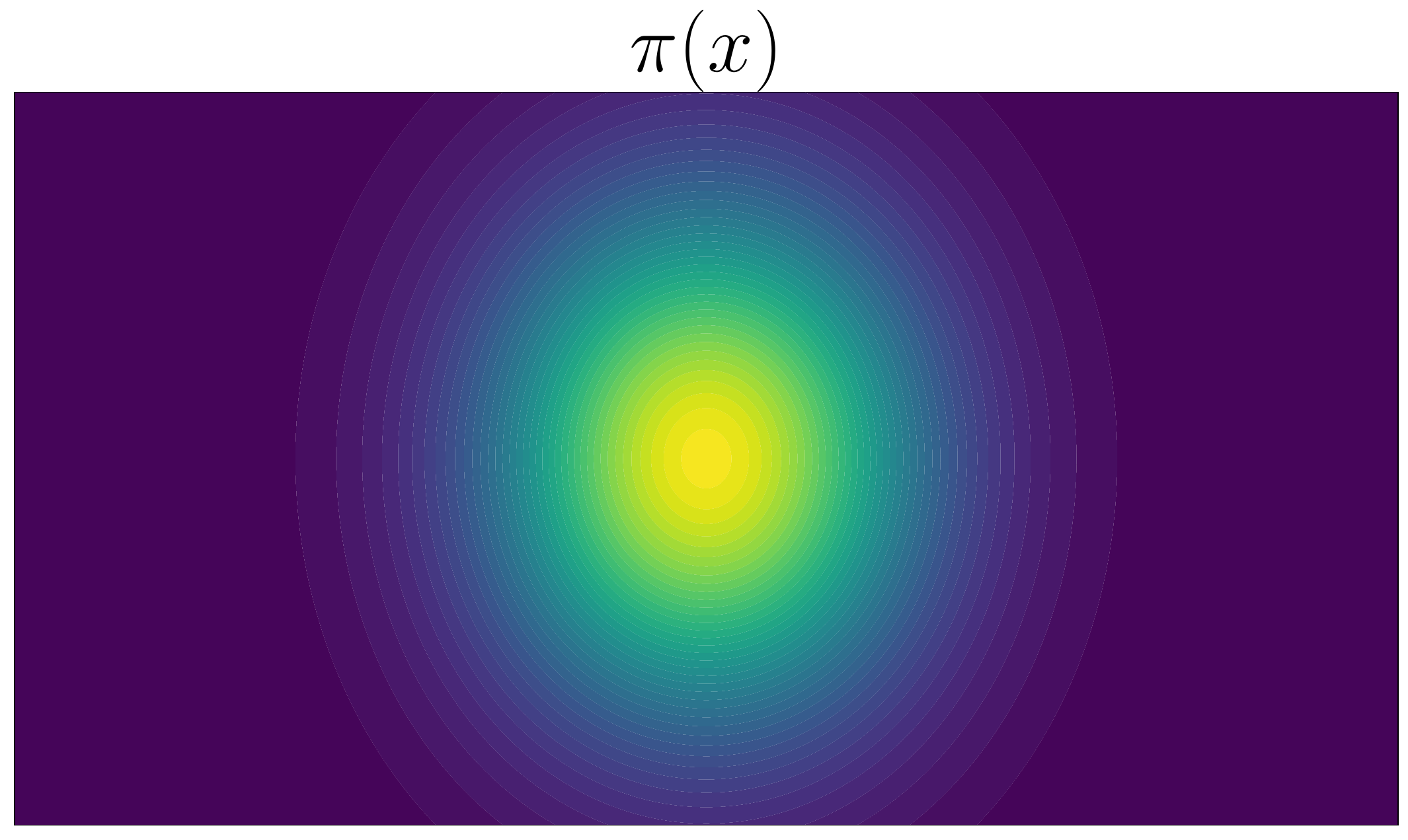}
        \label{fig:plot3}
    \end{subfigure}
    \caption{In this example, $q_{\text{SNIS}}^{\bigstar}$ is clearly different from both the optimal UIS proposal $\propto \pi(x) |\varphi(x)|$ and $\pi(x)$. }
    \label{fig:all_plots}
\end{figure*}

\begin{algorithm}
    \caption{ \ourmethod{} based on RWM  }
    \label{alg:simple-alg-2} 
        \textbf{Inputs. } 
        \begin{itemize}
            \item Iterations $T$, $J$ MCMC steps per iteration, burn-in $B < J$ in the first iteration.
            \item RWM symmetric proposal $q(\cdot | \cdot)$
            \item Initial estimate $\widehat{\mu}^{(0)}$ 
        \end{itemize}
        
         Choose initial state $x^{(0)}$.
        \begin{itemize}
        \item For $t=1,\dots,T$
        \begin{itemize} 
        \item For $j=1,\dots,J$
        \begin{itemize}
            \item If $j=1$, set $x_{\text{prev}}$ to $x^{(J)}$ from $t-1$ (or $x^{(0)}$). Otherwise, set $x_{\text{prev}}$ to $x^{(j-1)}$.
            \item Draw $y^{(j)} \iidsim q(\cdot | x_{\text{prev}})$
            \item Accept the new state $x^{(j)} = y^{(j)}$, with probability $ \alpha(x_{\text{prev}}, y^{(j)}) $ equal to
            \begin{align}\label{eq:acc_mh}
               \min \left (1, \frac{\widetilde{\pi}(y^{(j)}) |  \varphi(y^{(j)}) - \widehat{\mu}^{(t-1)}| }{\widetilde{\pi}(x_{\text{prev}})  | \varphi(x_{\text{prev}}) - \widehat{\mu}^{(t-1)}| } \right ) 
            \end{align}
            otherwise set $x^{(j)} = x_{\text{prev}}$. 
        \end{itemize}
        \item Compute the importance weights 
        \begin{equation}
           w^{(j)}:= w(x^{(j)}) = (|\varphi(x^{(j)}) - \widehat{\mu}^{(t-1)}|)^{-1} \nonumber
        \end{equation}
        \item   Compute current estimate of $\mu$, as 
        \begin{equation}\label{eq:current_est}
            \widehat{\mu}^{(t)} := \frac{ \sum_{i=1}^{P} w^{(i)} \varphi(x^{(i)})}{\sum_{i=1}^{P} w^{(i)} },
        \end{equation}
        where $P = 1$ if $j > 1$, otherwise $P = B$ as burn-in is only for $t=1$
    
        \end{itemize}
        \end{itemize}

        \textbf{Output.} The estimators $\{ \widehat{\mu}^{(t)}\}_{t=1}^{T}$. 
    \end{algorithm}

\subsection{Discussion and future directions}\label{sec:discussion}
In general, the densities $\pi$, $\pi |\varphi|$ and $q_{\text{SNIS}}^{\bigstar}$ can be significantly different, and this is a main conceptual motivation for targeting the optimal SNIS proposal.
Our framework and its potential extension can then obtain a significant advantage from optimizing the SNIS proposal, unlike most existing methods in the literature that explicitly or implicitly improve the UIS proposal, using instead SNIS for estimation.
 Other extensions of the proposed \ourmethod{} could include resampling and  tempering steps, which would connect the framework with sequential Monte Carlo (SMC) samplers \cite{del2006sequential}; the use of orthogonal chains \cite{martino2016orthogonal}; or the use of ergodic maps of recent algorithms combining MC and deterministic maps \cite{xu2023mixflows,thin2021neo}. Some initial theoretical guarantees of our framework could be borrowed using results from the RIS literature. Further, our Markov chain sequentially targets a sequence of approximations of the true optimal SNIS proposal. Thus, it connects with the noisy MCMC literature \cite{alquier2016noisy}, where some of the tools could be used to study the convergence of the chains. More specifically, we could view \cref{eq:acc_mh} as an approximation of a ``true'' acceptance ratio containing $\widetilde{\pi}(x) |\varphi(x) - \mu|$. 
 
\section{Illustrative example: Bayesian linear regression}\label{sec:experiments}

\begin{figure}[t]
    \centering
    \begin{subfigure}[b]{0.42\textwidth}
        \centering
\includegraphics[width=0.8\textwidth]{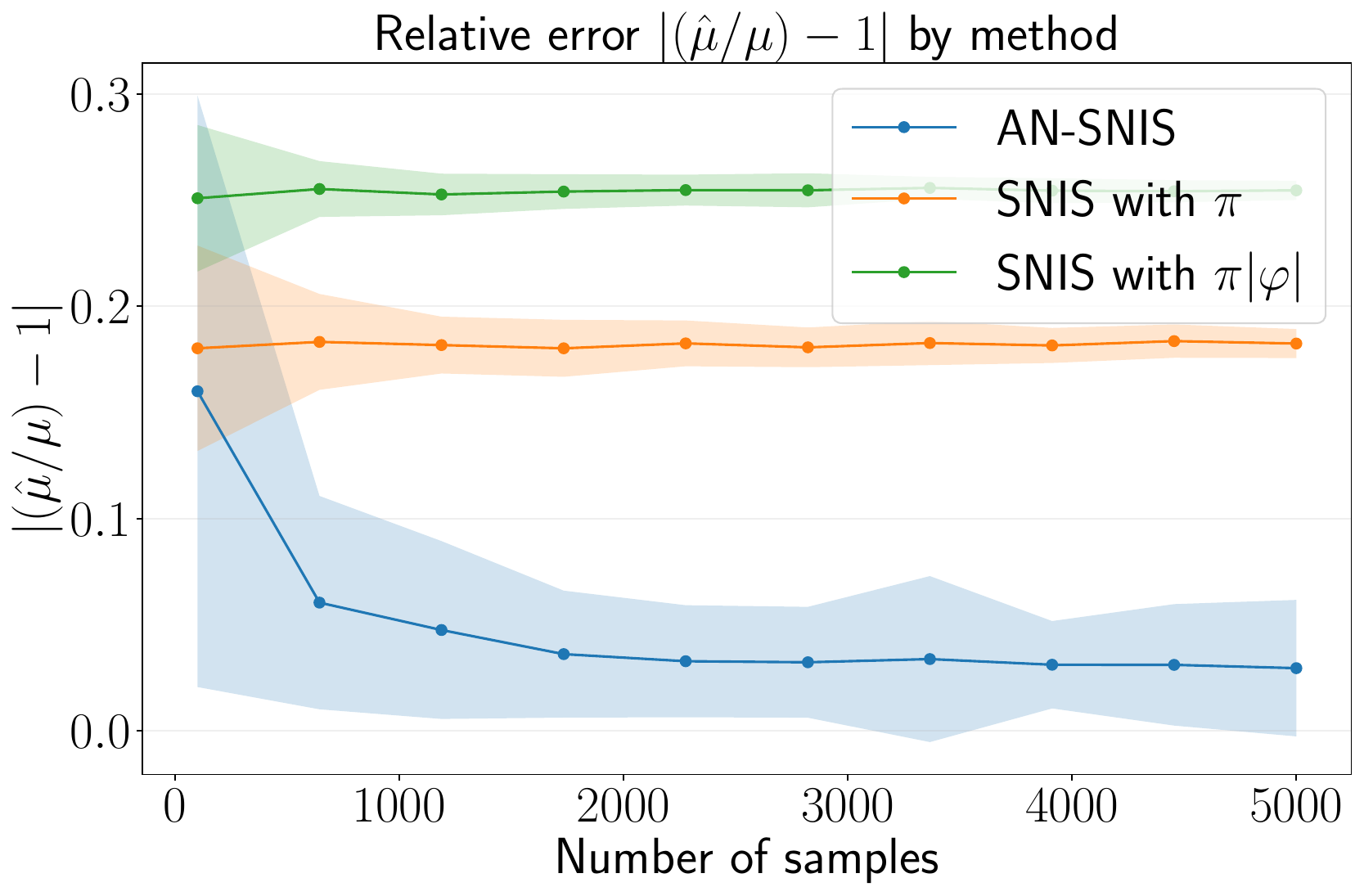}
        \label{fig:res1}
    \end{subfigure}
    \hfill
    \begin{subfigure}[b]{0.42\textwidth}
        \centering
        \includegraphics[width=0.8\textwidth]{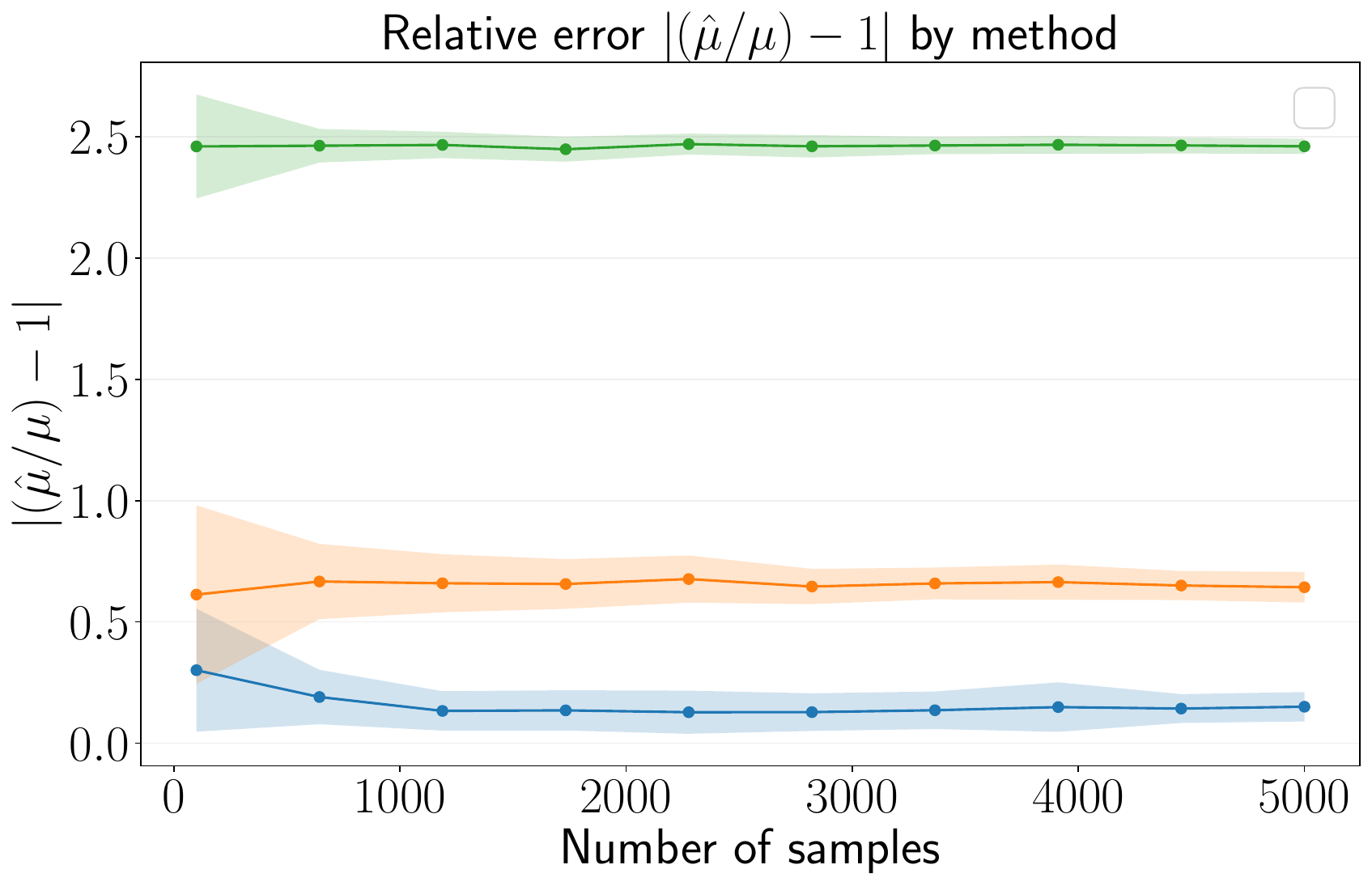}
        \label{fig:res2}
    \end{subfigure}
    \caption{Relative error of \ourmethod{} compared to SNIS with proposals proportional to $\pi |\varphi|$ and $\pi$. As $q_{\text{SNIS}^{\bigstar}}$ differs from both $\pi |\varphi|$ and $\pi$, we achieve significantly lower relative error, $|\widehat{\mu}/\mu -1|$. The bands display the mean $\pm$ one standard deviation obtained over $75$ replications.}
    \label{fig:results}
\end{figure}

We study the performance of \ourmethod{} in a Bayesian inference problem where $\mu$ is known in closed form, so we can evaluate all methods. 
We define the data  
$\mathcal{D} := \{ y_m \}_{m=1}^M$, with $y_m\in  \mathbb{R}$
and the target of Eq. \eqref{eq:integral} is the posterior PDF $\pi(x | \mathcal{D}) := Z_{\pi}^{-1} \cdot \widetilde{\pi}(x | \mathcal{D}) = Z_{\pi}^{-1} \cdot \prod_{m=1}^M p(y_m | x) \cdot \pi_0(x) $, $Z_{\pi}$ the normalizing constant and prior PDF $\pi_0(x)$. We are interested in approximating expectations under the posterior:
\begin{equation}\label{eq:integral_of_interest}
    \mu = \mathbb{E}_{\pi(x | \mathcal{D})}[\varphi(x)] = \int \varphi(x) \pi(x | \mathcal{D}) d x , ~~ x \in \mathbb{R}^D .
\end{equation}
In Bayesian linear regression (BLR) with a Gaussian prior, the posterior is also Gaussian and can be calculated analytically. 
A function of particular interest, for instance in Bayesian cross validation \cite{vehtari2017practical}, is $\varphi(x) = p(y_{M+1}|x)$, where $y_{M+1}$ is a \emph{test} sample.
Even in this simple setting, where $\pi$ is Gaussian and so is the optimal UIS proposal for \cref{eq:integral_of_interest}, the optimal SNIS proposal is very different from both functions (even multimodal). We visualize this in \cref{fig:all_plots}. 

The aim in our experiments is to demonstrate that it can be beneficial to attempt to use a proposal that is close to $q_{\text{SNIS}}^{\bigstar}$, and that \ourmethod{} successfully does so. We will compare our \ourmethod{}, which adaptively targets the optimal SNIS proposal $\propto \pi(x) |\varphi(x) - \mu|$, with the following options
\begin{enumerate}
    \item \texttt{SNIS-UISOPT}: SNIS estimator using $\pi(x)$ as proposal
    \item \texttt{SNIS-TARGET}: SNIS estimator using $\propto \pi(x) |\varphi(x)|$ as proposal, i.e., the optimal UIS proposal .
\end{enumerate}
For this application, both $\pi$ and $\pi |\varphi|$ are Gaussian. For fair comparison, for all methods we sample using MCMC with the same settings. Despite $q_{\text{SNIS}}^{\bigstar}$ being potentially more complex to sample from, our method achieves better performance with the same MCMC setup. We measure performance via relative error $| (\widehat{\mu}/\mu ) - 1|$ \cite{chatterjee2018sample}, and use $1000 \cdot D$ burn-in samples for all algorithms.
\textbf{Fair sample allocation.} For 2$D$ experiments, we used 1000 additional samples for $\widehat{\mu}^{(0)}$ estimation. This initial estimate is obtained with UIS, with a proposal being a perturbed version of $|\varphi| \pi$, as: $\sigma_{d}^2 = \sigma_{d, |\varphi| \pi}^2 + \varepsilon$, $\mu_d = \mu_{d, |\varphi| \pi} + \epsilon$ where $\varepsilon = 0.05 / D$. For the varying dimension experiments, we ensured a fairer comparison by subtracting $N_{\text{prelim}}=0.1N$ from \ourmethod{}'s budget.
\textbf{MCMC details.} We used RWM with a Gaussian proposal and stepsize following optimal scaling guidelines as our targets are factorized (see, e.g., \cite{titsias2023optimal}).

\textbf{Example 1.} This is a $2D$ example whose distributions are visualized in \cref{fig:all_plots}. Here $D=2$, $\pi(x | \mathcal{D})$ has variances $\sigma_{1,\pi}^2,\sigma_{2,\pi}^2 = 12/1000, 6/100$; $\varphi(x)$ has variances $\sigma_{1,\varphi}^{2},\sigma_{2,\varphi}^{2} = 12/100, 6/100$. \textbf{Example 2.} Here, also $D=2$, $\pi(x | \mathcal{D})$ has variances $\sigma_{1,\pi}^2,\sigma_{2,\pi}^2 = 5/100, 1/100$; $\varphi(x)$ has variances $\sigma_{1,\varphi}^{2},\sigma_{2,\varphi}^{2} = 5/1000, 5/1000$. We always have zero means.

The results of both examples are shown in \cref{fig:results}. We found similar results for other stepsizes and choices where the SNIS optimal proposal differs significantly from $\pi$ and $\pi |\varphi|$. 

\begin{figure}[t]
    \centering
    \begin{subfigure}[b]{0.22\textwidth}
        \centering
        \includegraphics[width=\textwidth]{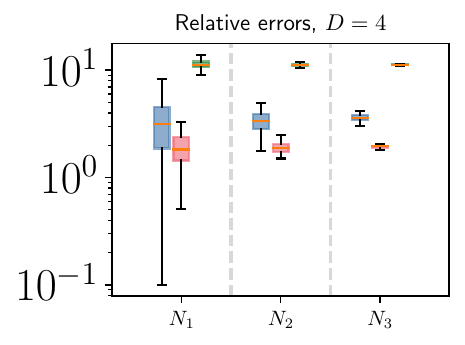}
    \end{subfigure}%
    \begin{subfigure}[b]{0.22\textwidth}
        \centering
        \includegraphics[width=\textwidth]{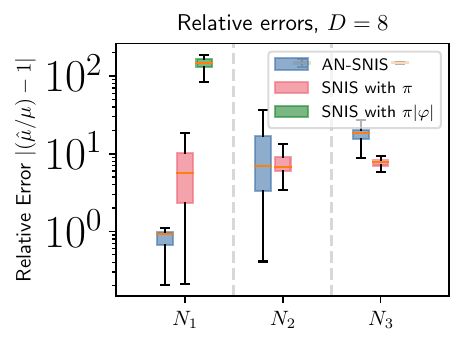}
    \end{subfigure}
    
    \begin{subfigure}[b]{0.22\textwidth}
        \centering
        \includegraphics[width=\textwidth]{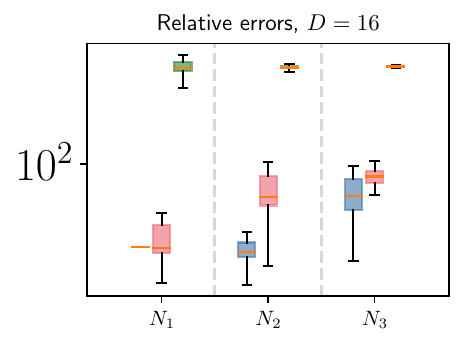}
    \end{subfigure}%
    \begin{subfigure}[b]{0.22\textwidth}
        \centering
        \includegraphics[width=\textwidth]{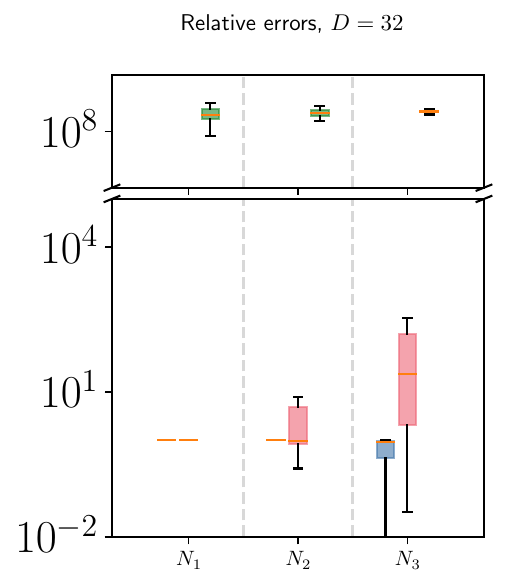}
    \end{subfigure}
    \caption{\footnotesize Relative error comparison across different dimensions ($D=4,8,16,32$), results over $50$ replications.}
    \label{fig:dim_grid}
\end{figure}

\textbf{Varying dimensions.} In \cref{fig:dim_grid}, we show results for varying dimensions. We used sample sizes $N_1 = 100 \cdot D^{3/2}, N_2 = 1000 \cdot D^{3/2}, N_3 = 10000 \cdot D^{3/2}$, i.e., scaling with dimension. Here, $\pi$ has isotropic covariance with $\sigma_{\pi} = (0.5/D) \cdot I_{D}$ and $\varphi$ has covariance $\sigma_{\varphi} = (0.1/D) \cdot I_{D}$, following intuition from $2D$ examples to have larger $\pi$ variance than $\varphi$. Other values for $\sigma_{\varphi}, \sigma_{\pi}$ either lead to similar results or all algorithms struggle equally. More testing, especially on more complex examples, is beyond the current scope.
\section{Conclusion}\label{sec:conclusions}
We have introduced \ourmethod{}, the first AIS framework specifically designed for the SNIS estimator. Unlike most AIS samplers, \ourmethod{} does not restrict the proposal in a simple parametric family. We have highlighted connections with the ratio IS literature, paving the way to even better IS samplers for the SNIS estimator. We have also outlined new interesting theoretical directions connecting with other methods, including noisy MCMC. Finally, we have demonstrated the improved performance brought by \ourmethod{} in an illustrative example where the optimal SNIS proposal is very different from other standard choices.

\clearpage 

\bibliographystyle{IEEEtran}
\bibliography{refs.bib}

\end{document}